\documentclass[12pt,preprint,usenatbib]{aastex}

\usepackage{amssymb}
\usepackage{times}
\usepackage{pifont} 

\newcommand{\teff}{$T_{\mathrm{eff}}$}
\newcommand{\muhz}{$\mu$Hz}
\newcommand{\numax}{$\nu_{\mathrm{max}}$}
\newcommand{\dnu}{$\Delta\nu$}

\shorttitle{Solar-like oscillations in NGC~6819 from {\it Kepler} photometry}
\shortauthors{Stello et al.}

\begin{document}

\title{Detection of solar-like oscillations from {\it Kepler} photometry of the open
cluster NGC~6819}  

\author{
Dennis~Stello,\altaffilmark{1} 
Sarbani~Basu,\altaffilmark{2} 
Hans~Bruntt,\altaffilmark{3} 
Beno\^it~Mosser,\altaffilmark{3} 
Ian~R.~Stevens,\altaffilmark{4} 
Timothy~M.~Brown,\altaffilmark{5} 
J{\o}rgen~Christensen-Dalsgaard,\altaffilmark{6} 
Ronald~L.~Gilliland,\altaffilmark{7} 
Hans~Kjeldsen,\altaffilmark{6} 
Torben~Arentoft,\altaffilmark{6} 
J\'er\^ome~Ballot,\altaffilmark{8} 
Caroline~Barban,\altaffilmark{3} 
Timothy~R.~Bedding,\altaffilmark{1} 
William~J.~Chaplin,\altaffilmark{4} 
Yvonne~P.~Elsworth,\altaffilmark{4} 
Rafael~A.~Garc{\'\i}a,\altaffilmark{9} 
Marie-Jo~Goupil,\altaffilmark{3} 
Saskia~Hekker,\altaffilmark{4} 
Daniel~Huber,\altaffilmark{1} 
Savita~Mathur,\altaffilmark{10} 
S{\o}ren~Meibom,\altaffilmark{11} 
Reza~Samadi,\altaffilmark{3} 
Vinothini~Sangaralingam,\altaffilmark{4} 
Charles~S.~Baldner,\altaffilmark{2} 
Kevin~Belkacem,\altaffilmark{12} 
Katia~Biazzo,\altaffilmark{13} 
Karsten~Brogaard,\altaffilmark{6} 
Juan~Carlos~Su\'arez,\altaffilmark{14} 
Francesca~D'Antona,\altaffilmark{15} 
Pierre~Demarque,\altaffilmark{2} 
Lisa~Esch,\altaffilmark{2} 
Ning~Gai,\altaffilmark{2,16} 
Frank~Grundahl,\altaffilmark{6} 
Yveline~Lebreton,\altaffilmark{17} 
Biwei~Jiang,\altaffilmark{16} 
Nada~Jevtic,\altaffilmark{18} 
Christoffer~Karoff,\altaffilmark{4} 
Andrea~Miglio,\altaffilmark{12} 
Joanna~Molenda-\.Zakowicz,\altaffilmark{19} 
Josefina~Montalb\'an,\altaffilmark{12} 
Arlette~Noels,\altaffilmark{12} 
Teodoro~Roca~Cort\'es,\altaffilmark{20,21} 
Ian~W.~Roxburgh,\altaffilmark{22} 
Aldo~M.~Serenelli,\altaffilmark{23} 
Victor~Silva~Aguirre,\altaffilmark{23} 
Christiaan~Sterken,\altaffilmark{24}
Peter~Stine,\altaffilmark{18}
Robert~Szab\'o,\altaffilmark{25}
Achim~Weiss,\altaffilmark{23}
William~J.~Borucki,\altaffilmark{26}
David~Koch,\altaffilmark{26}
Jon~M.~Jenkins\altaffilmark{27}
}
\altaffiltext{1}{Sydney Institute for Astronomy (SIfA), School of Physics, University of Sydney, NSW 2006, Australia}
\altaffiltext{2}{Department of Astronomy, Yale University, P.O. Box 208101, New Haven, CT 06520-8101}
\altaffiltext{3}{LESIA, CNRS, Universit\'e Pierre et Marie Curie, Universit\'e Denis Diderot, Observatoire de Paris, 92195 Meudon, France}
\altaffiltext{4}{School of Physics and Astronomy, University of Birmingham, Edgbaston, Birmingham B15 2TT, UK}
\altaffiltext{5}{Las Cumbres Observatory Global Telescope, Goleta, CA 93117, USA}
\altaffiltext{6}{Department of Physics and Astronomy, Aarhus University, 8000 Aarhus C, Denmark}
\altaffiltext{7}{Space Telescope Science Institute, 3700 San Martin Drive, Baltimore, Maryland 21218, USA}
\altaffiltext{8}{Laboratoire d'Astrophysique de Toulouse-Tarbes, Universit\'e de Toulouse, CNRS,14 av E. Belin, 31400 Toulouse, France}
\altaffiltext{9}{Laboratoire AIM, CEA/DSM-CNRS, Universit\'e Paris 7 Diderot, IRFU/SAp, Centre de Saclay, 91191, Gif-sur-Yvette, France}
\altaffiltext{10}{Indian Institute of Astrophysics, Koramangala, Bangalore 560034, India}
\altaffiltext{11}{Harvard-Smithsonian Center for Astrophysics, 60 Garden Street, Cambridge, MA, 02138, USA}
\altaffiltext{12}{Institut d'Astrophysique et de G\'eophysique de l'Universit\'e de Li\`ege, 17 All\'ee du 6 Ao\^ut, B-4000 Li\`ege, Belgium}
\altaffiltext{13}{Arcetri Astrophysical Observatory, Largo E. Fermi 5, 50125, Firenze, Italy}
\altaffiltext{14}{Instituto de Astrof\'{\i}sica de Andaluc\'{\i}a (CSIC),Dept. Stellar Physics, C.P. 3004, Granada, Spain}
\altaffiltext{15}{INAF - Osservatorio di Roma, via di Frascati 33, I-00040, Monteporzio, Italy}
\altaffiltext{16}{Department~of~Astronomy,~Beijing~Normal~University, Beijing 100875, China}
\altaffiltext{17}{GEPI, Observatoire de Paris, CNRS, Universit\'e  Paris Diderot, 5 Place Jules Janssen, 92195 Meudon, France}
\altaffiltext{18}{Department of Physics \& Engineering Technology, Bloomsburg University, 400 East Second St, Bloomsburg PA 17815, USA}
\altaffiltext{19}{Astronomical Institute, University of Wroc\l aw, ul.\ Kopernika 11, 51-622 Wroc\l aw, Poland}
\altaffiltext{20}{Departmento de Astrof\'{i}ca, Universidad de La Laguna, 38207 La Laguna, Tenerife, Spain}
\altaffiltext{21}{Instituto de Astrof\'{i}ca de Canarias, 38205 La Laguna, Tenerife, Spain}
\altaffiltext{22}{Queen Mary University of London, Mile End Road, London E1 4NS, UK}
\altaffiltext{23}{Max Planck Institute for Astrophysics, Karl Schwarzschild Str. 1, Garching bei M\"{u}nchen, D-85741, Germany}
\altaffiltext{24}{Vrije Universiteit Brussel, Pleinlaan 2, B-1050 Brussels, Belgium}
\altaffiltext{25}{Konkoly Observatory, H-1525 Budapest, P.O. Box 67, Hungary}
\altaffiltext{26}{NASA Ames Research Center, MS 244-30, Moffat Field, CA 94035, USA}
\altaffiltext{27}{SETI Institute/NASA Ames Research Center, MS 244-30, Moffat Field, CA 94035, USA}

\clearpage

\begin{abstract}
Asteroseismology of stars in clusters has been a long-sought goal because
the assumption of a common age, distance and initial chemical composition 
allows strong tests of the theory of stellar evolution.  We report results from
the first 34 days of science data from the {\it Kepler Mission} for the open
cluster NGC~6819 --- one of four clusters in the field of view.  We obtain
the first clear detections of solar-like oscillations in the cluster red giants and
are able to measure the large frequency separation, $\Delta\nu$, and the
frequency of maximum oscillation power, $\nu_{\rm max}$. 
We find that the asteroseismic parameters allow us to test
cluster-membership of the stars, and even with the limited seismic
data in hand, we can already identify four possible non-members despite
their having a better than 80\% membership probability from radial velocity
measurements.  
We are also able to determine the oscillation amplitudes for stars that span about
two orders of magnitude in luminosity and find good agreement with the
prediction that oscillation amplitudes scale as the luminosity to the power
of 0.7.
These early results demonstrate the unique potential of asteroseismology
of the stellar clusters observed by {\it Kepler}.

\end{abstract}


\clearpage

\section{Introduction} 
Open clusters provide unique opportunities in astrophysics.  
Stars in open clusters are believed to be formed from the same cloud of gas
at roughly the same time.  
The fewer free parameters available to model cluster stars make them interesting targets to
analyze as a uniform ensemble, especially for asteroseismic studies.
 
Asteroseismology is an elegant tool based on the simple principle that the
frequency of a standing acoustic wave inside a star depends on the sound
speed, which in turn depends on the physical properties of the interior.  
This technique applied to the Sun (helioseismology) has provided extremely
detailed knowledge about the physics that 
governs the solar interior, \citep[e.g.,][]{Dalsgaard02}.
All cool stars are expected to exhibit solar-like oscillations of
standing acoustic waves -- called p modes -- that are stochastically driven by
surface convection. 
Using asteroseismology to probe the interiors of
cool stars in clusters, therefore, holds promise of rewarding scientific return
\citep{GoughNovotny93,BrownGilliland94}.
This potential has resulted in
several attempts to detect solar-like oscillations in clusters using time-series
photometry. These attempts were often aimed at red giants, since their oscillation
amplitudes are expected to be larger than those of main-sequence or
subgiant stars due to more vigorous surface convection.  Despite these
attempts, only marginal detections have 
been attained so far, limited either by the length of the time series
usually achievable through observations with the {\em Hubble Space Telescope}
\citep{EdmondsGilliland96,StelloGilliland09} or by the difficulty in
attaining high precision from ground-based campaigns
\citep[e.g.,][]{Gilliland93,Stello07,Frandsen07}.

In this Letter we report clear detections of solar-like oscillations 
in red-giant stars in the open cluster NGC~6819 using photometry from
NASA's {\it Kepler Mission} \citep{Borucki09}. 
This cluster, one of four in the {\it Kepler} field, is about 2.5~Gyr old. It is at a distance of
2.3~kpc, and has a metallicity of [Fe/H]~$\sim -0.05$
\citep[see][~and references herein]{Hole09}.

\section{Observations and data reduction}\label{observations}
\begin{figure}
\plotone{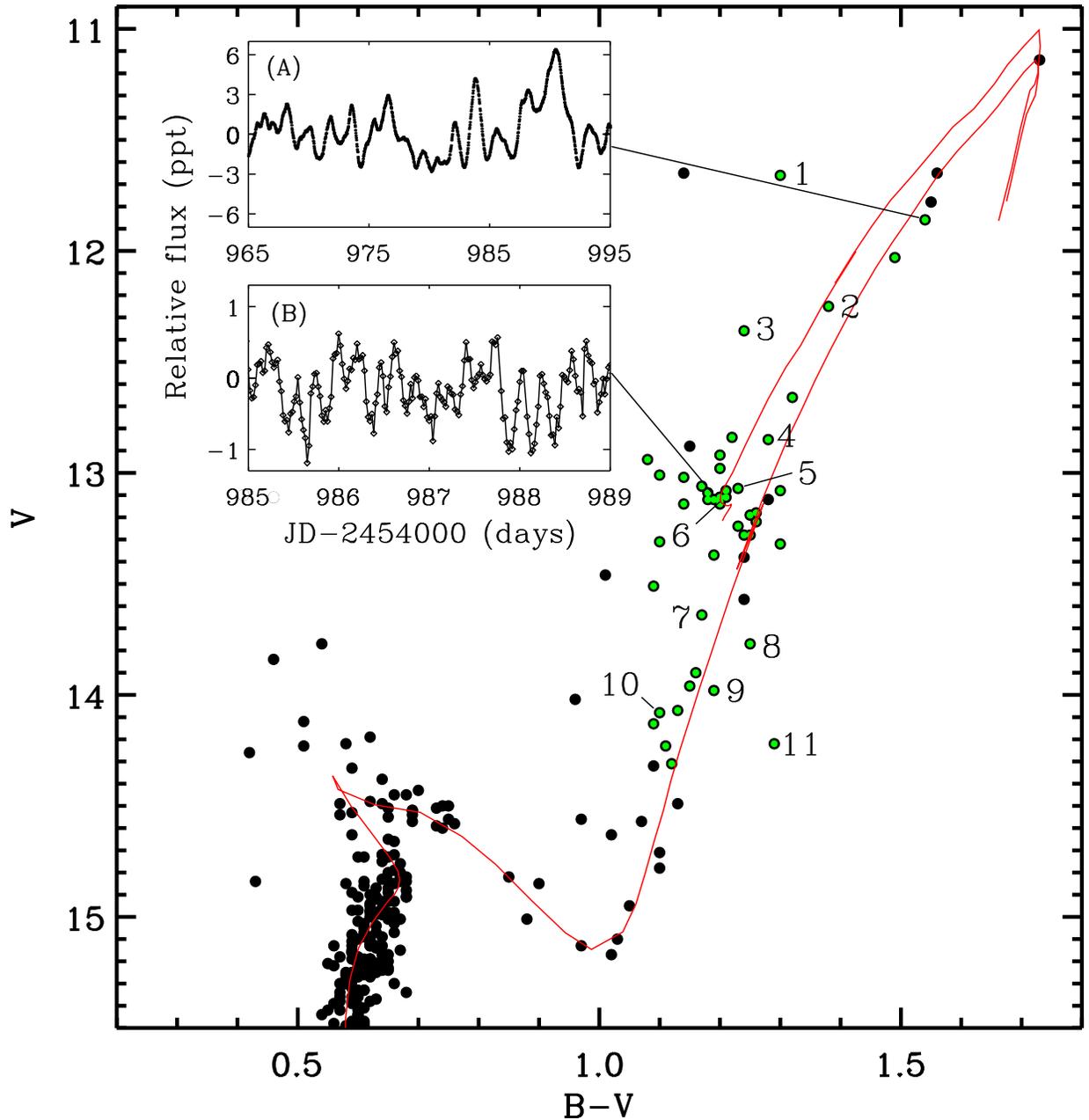}
\caption{
  Color-magnitude diagram of NGC~6819.  Plotted stars have membership
  probability $P_{\mathrm{RV}}>80$\% as determined by \citet{Hole09}. Photometric
  indices are from the same source.  The isochrone is from
  \citet{Marigo08} (Age=2.4 Gyr, Z=0.019, modified for the adopted 
   reddening of 0.1mag).  Color-coded stars have been
  analyzed, and the annotated numbers refer to the legend in panels of Figure~\ref{f2} and
  star numbers in Figure~\ref{f3} (see also
  Table~\ref{tab1}).  Insets show light curves in parts per thousand of two red giants
  oscillating on different timescales. The variations of the light 
  curves in  Panel A and B 
  are dominated by the stellar oscillations with periods of a few days
   and of about six hours, respectively.
\label{f1}} 
\end{figure}

The data were obtained between 2009 May 12 and
June 14, i.e., the first 34 days of continuous science
observations by {\it Kepler}  (Q1 phase).
The spacecraft's long-cadence mode 
($\Delta t \simeq 30\,$minutes) used in this investigation provided a 
total of 1639 data points in the time series of each observed star.
For this Letter we selected 47 stars in the field of the open
cluster NGC~6819 with  membership probability $P_{\mathrm{RV}}>80$\% 
from radial velocity measurements \citep{Hole09}.  Figure~\ref{f1} shows the
color-magnitude diagram (CMD) of the cluster with the selected stars
indicated by green symbols. The eleven annotated stars form a representative
subset, which we will use to illustrate our analyses in 
Sections~\ref{extract} and~\ref{membership}.  We selected the stars in this
subset to cover the same brightness range as our full sample, while giving
high weight to stars that appear to be photometric non-members (i.e., stars
located far from the isochrone in the CMD). 
Data for each target were checked carefully to ensure that the time-series
photometry was not contaminated significantly by other stars in the field,
which could otherwise complicate the interpretation of the oscillation
signal.  

Fourteen data points affected by the momentum dumping of the spacecraft 
were removed from the time series of each star.  In addition, we removed
points that showed a point-to-point deviation greater than 
$4\sigma$, where $\sigma$ is the local rms of the point-to-point scatter
within a 24 hour window.  This process removed on average one data-point
per time series.  
Finally, we removed a linear trend from each time series and then
calculated the discrete Fourier transform. The
Fourier spectra at high frequency have mean levels below 5 parts per
million (ppm) in amplitude, allowing us to search for low-amplitude
solar-like oscillations.

\section{Extraction of asteroseismic parameters}\label{extract}
\begin{figure}
\epsscale{0.8}
\plotone{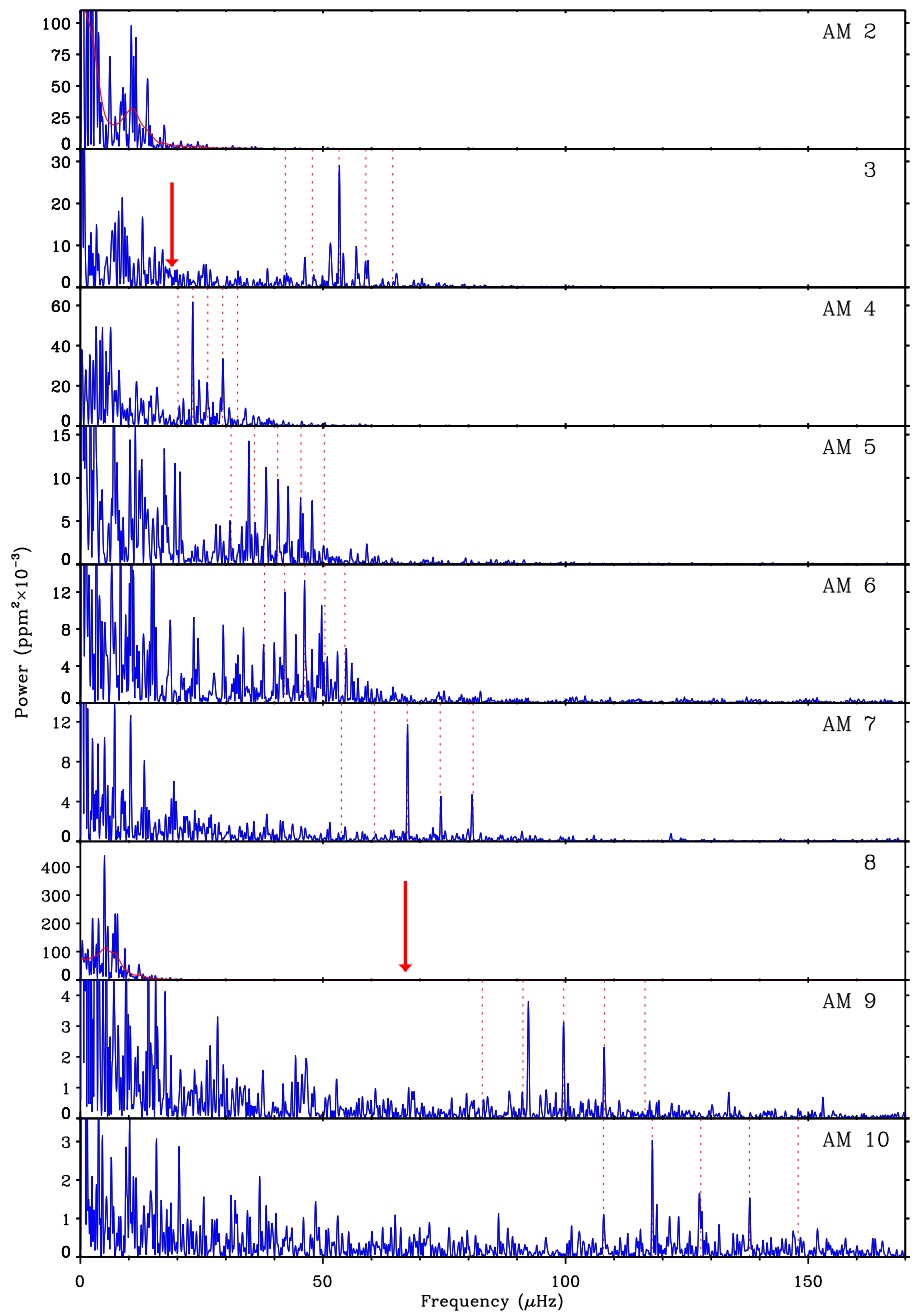}
\caption{
  Fourier spectra of a representative set of red giants along the cluster
  sequence sorted by apparent magnitude. Annotated numbers in each
  panel refer to the star identification (see Fig.~\ref{f1} and
  Table~\ref{tab1}). 
  `AM' indicates that the star is an asteroseismic 
  member. Red solid curves show the smoothed spectrum for stars with
  \numax\ $< 20\,$\muhz.  To guide the eye, we have plotted dotted lines to
  indicate the measured average large frequency separation. The central
  dotted  line is centered on the highest oscillation peaks near
  \numax. Note that since $\Delta\nu$ is generally frequency dependent,
  only the central dotted line is expected to line up with a peak in the
  oscillation spectrum.  The red arrows indicate the position of the
  expected \numax\ (see Eq.~\ref{numaxsc}) for stars where the observed
  value does not agree with the expectations for this cluster (see
  Section~\ref{membership}). 
  \label{f2}} 
\end{figure}

Figure~\ref{f2} shows the Fourier spectra (in power) of 9 stars from
our subset. These range from the lower red-giant branch to the tip of the branch
(see Figure~\ref{f1}). 
The stars are sorted by apparent magnitude, which for a cluster is
indicative of luminosity, with brightest at the top.  
Note that the red giants in NGC~6819 are significantly fainter ($12\lesssim
V \lesssim 14$) than the sample of {\em Kepler} field red giants ($8\lesssim
V \lesssim 12$) studied by \citet{Bedding10}. Nevertheless, it is clear
from Figure~\ref{f2} that we can detect 
oscillations for stars that span about two orders of magnitude in
luminosity along the cluster sequence. 

We used four different pipelines
\citep{Hekker09a,Huber09,Mathur09a,MosserAppourchaux09} to 
extract the average frequency separation between 
modes of the same degree (the so-called large frequency separation, \dnu).
We have also obtained the frequency of maximum oscillation
power, \numax, and the oscillation amplitude.
The measured values of \dnu\ are indicated by vertical dotted
lines in Figure~\ref{f2} centered on the highest oscillation peaks near
\numax.
While the stars in Figure~\ref{f2}, particularly in the lower panels,
show the regular series of peaks expected for solar-like oscillations, the
limited length of the time-series data does not allow such structure to be
clearly resolved for the most luminous stars in our sample --- those with
\numax\ $\lesssim 20\,$\muhz. 
We do, however, see humps of excess power in the Fourier spectra (see
Figure~\ref{f2} star no. 2 and 8) with
\numax\ and amplitude in mutual agreement with oscillations.
With longer time series we expect more firm results for these 
high-luminosity giants.

\section{Cluster membership from asteroseismology}\label{membership}
It is immediately clear from Figure~\ref{f2} that not all stars 
follow the expected trend of increasing \numax\ with decreasing apparent
magnitude, suggesting that some of the stars might be intrinsically
brighter or fainter than expected.
Since oscillations in a star only depend on the physical properties of the star, 
we can use asteroseismology to judge whether or not a star is
likely to be a cluster member independently of its distance and 
of interstellar absorption and reddening. 
For cool stars, \numax\ scales with the acoustic 
cut-off frequency, and it is well established that we can estimate 
\numax\ by scaling from the solar value
\citep{Brown91,KjeldsenBedding95}:   
\begin{eqnarray}
\frac{\nu_{\mathrm{max}}}{\nu_{\mathrm{max,\odot}}}
                  =\frac{M/M_\odot (T_{\mathrm{eff}}/T_{\mathrm{eff,\odot}})^{3.5}}{L/L_\odot},
\label{numaxsc}
\end{eqnarray}
where $\nu_{\mathrm{max,\odot}}=3100\,\mu$Hz.  The accuracy of such 
estimates is good to within 5\% \citep{Stello09a} assuming we have
good estimates of the stellar parameters $M$, $L$, and \teff.

\begin{figure}
\epsscale{1}
\plotone{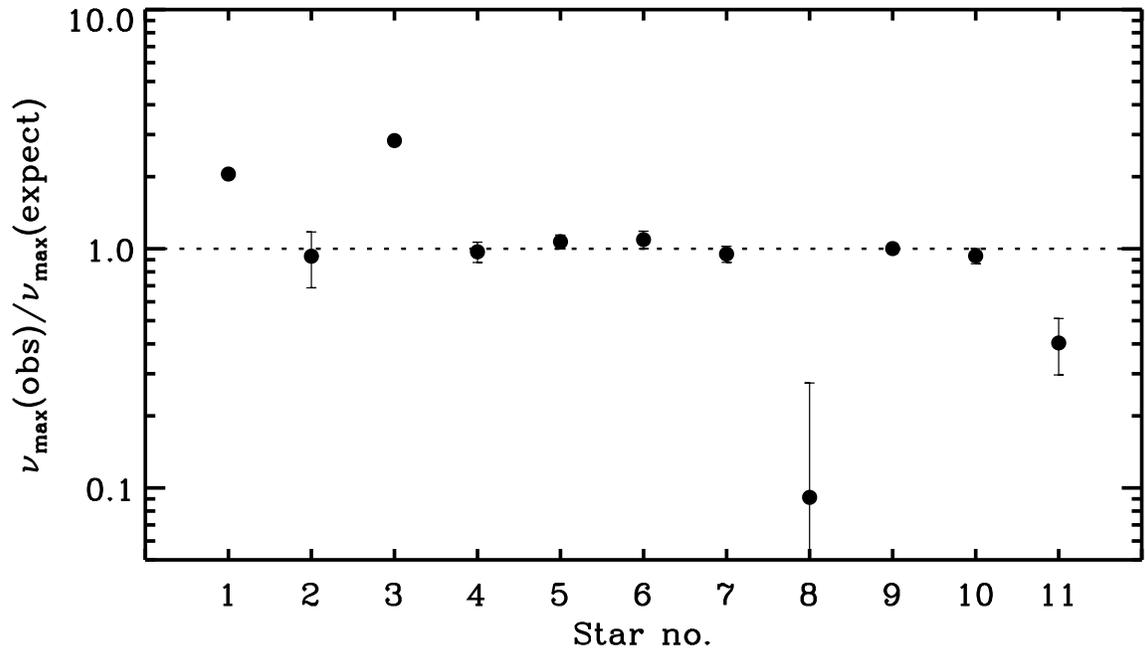}
\caption{
  Ratio of observed and expected \numax. 1-$\sigma$ errorbars indicate the
  uncertainty on \numax(obs). Stars clearly
  above or below the dotted line are either not cluster members or 
  members whose evolution have not followed the standard scenario.
\label{f3}} 
\end{figure}

In the following we assume the idealistic scenario where all cluster members
follow standard stellar evolution described by the isochrone. 
Stellar mass along the red giant branch of the cluster isochrone varies by
less than 1\%. 
The variation is less than 5\% even if we also consider the
asymptotic giant branch.  For simplicity, we therefore adopt a mass of
$1.55M_\odot$ for all stars, which is representative for the isochrone from
\citet{Marigo08} (Figure~\ref{f1}) and a similar
isochrone by \citet{VandenBerg06}. 
Neglecting binarity (see Table~\ref{tab1}), we derive the luminosity
of each star in our subset from its $V$-band apparent magnitude,
adopting reddening and distance modulus of $E(B-V)=0.1$ and $(M-m)_V=12.3$,
respectively (obtained from simple isochrone fitting, see \citealt{Hole09}). 
We used the calibration of \citet{Flower96} to convert the stellar $(B-V)_0$
color to \teff.  Bolometric corrections were also taken from
\citet{Flower96}.  The derived quantities were then used to estimate
\numax\ for each star (Eq.~\ref{numaxsc}),  and compared
with the observed value (see Figure~\ref{f3}). 

Figure~\ref{f3} shows four obvious outliers (no. 1, 3, 8 and 11), three of which are also
outliers in the CMD (no. 1, 3, and 11). 
For the rest of the stars we see good agreement between the expected
and observed value, indicating that the uncertainty on the \numax\ estimates
are relatively small.
Since the variations in mass and
effective temperature among the cluster giant stars are 
small, deviations from the dotted line must be caused by an incorrect 
estimate of the luminosity. 
This implies that the luminosities of stars
falling significantly above or below the line have been over- or
underestimated, respectively.  
The simplest interpretation is that these outliers are fore- or
background stars, and hence not members of the cluster.
To explain the differences between the observed and expected value of
\numax\  would require the deviant stars to have $V$ errors of more than 1
magnitude, and in some cases $B-V$ errors of about 0.2 magnitude if they
were cluster members.  
Binarity may explain deviations above the dotted line, but only by up to a
factor of two in $L$ (and hence, in the ratio of the observed to expected \numax).  The
deviation of only one star 
(no.1) could potentially be explained this way.  However, that would be
in disagreement with its single-star classification from multi-epoch
radial velocity measurements, assuming it is not a binary viewed pole-on
(see Table~\ref{tab1}).  
Hence, under the assumption of a standard
stellar evolution, the most likely explanation for all four outliers in
Figure~\ref{f3} is therefore that these stars are not cluster members. 
This conclusion is, however, in disagreement with their high membership
probability from measurements of radial velocity \citep{Hole09} and
proper motion \citep{Sanders72} (see Table~\ref{tab1}). 
Another interesting possibility is that the anomalous pulsation properties
might be explained by more exotic stellar evolution scenarios than is generally
anticipated for open-cluster stars.

\section{Asteroseismic ``color-magnitude diagrams''}
It is clear from  Figure~\ref{f2} that the amplitudes of the 
oscillations increase with luminosity for the seismically determined
cluster members.
Based on calculations by \cite{DalsgaardFrandsen83},
\citet{KjeldsenBedding95} have suggested that the photometric oscillation
amplitude of p modes scale as
$(L/M)^{s}\,$\teff$^{-2}$, with $s=1$ (the velocity amplitudes, meanwhile,
would scale as $(L/M)^{s}$). This was revised by \citet{Samadi07}
to $s=0.7$ based on models of main sequence stars. 
Taking advantage of the fewer free parameters within this ensemble of
stars, our observations allow us to make some progress towards 
extrapolating this scaling to red giants and determining the value of $s$.

\begin{figure}
\epsscale{1}
\plotone{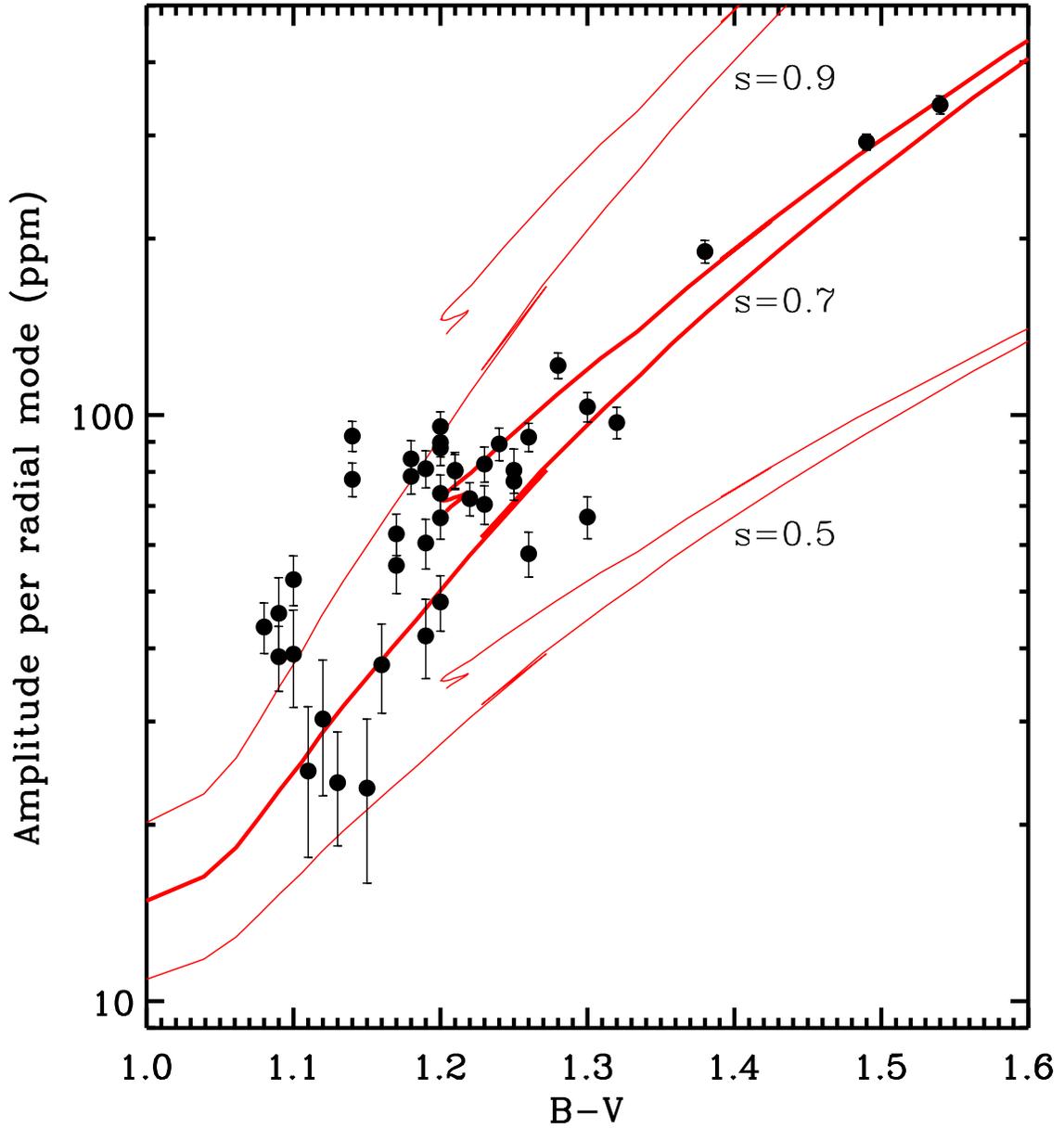}
\caption{
  Amplitude color diagram of red giant stars in NGC~6819 with the
  \citet{Marigo08} isochrone overlaid with three values of $s$ in the
  amplitude scaling relation: $(L/M)^{s}\,$\teff$^{-2}$. The solar value
  used in this scaling is $4.7\,$ppm  \citep{KjeldsenBedding95}. 
\label{f4}} 
\end{figure}

In Figure~\ref{f4} we introduce a new type of diagram that is similar to a CMD,
but with magnitude replaced by an asteroseismic parameter -- in this case,
the measured oscillation amplitude. Amplitudes were estimated for all stars
in our sample (except for the four outliers) using methods similar to that of
\citet{Kjeldsen07} \citep[see also][]{Michel08}, which assume that the
relative power between radial and non-radial modes is the same as in the Sun. 
This diagram confirms the relationship between 
amplitude and luminosity.  Despite a large 
scatter, which is not surprising from this relatively short time series, 
we see that $s=0.7$ provides a much better match than $s=1.0$. 
Once verified with more data, this relation will allow the use of
the measured amplitude as an additional asteroseismic diagnostic for 
testing cluster membership and for isochrone fitting in general.  
We note that the other clusters observed by {\em Kepler} have different
metallicities than NGC~6819, which will allow future investigation on the
metallicity dependence of the oscillation amplitudes.

We expect to obtain less scatter in
the asteroseismic measurements when longer time series become
available. That will enable us to expand classical isochrone fitting
techniques to include diagrams like this, where amplitude could also be
replaced by \numax\ or \dnu. 
In particular, we should be able to determine the absolute radii aided by
\dnu\ of the red giant branch stars, which would be an important calibrator for
theoretical isochrones. 
Additionally, the distributions of the asteroseismic
parameters -- such as \numax\ -- can potentially be used to test stellar
population synthesis models \citep{Hekker09,Miglio09}. 
Applying this
approach to clusters could lead to further progress in understanding
of physical processes such as 
mass loss during the red-giant phase \citep[see e.g.,][]{Miglio09a}.
Note that a few clear outliers are indicative of
non-membership or exotic stellar evolution, as a result of factors such
as stellar collisions or heavy mass loss, while a general deviation from
the theoretical predictions by a large group of stars would  suggest that the
standard theory may need revision.

Finally, we note that NGC~6819 and another {\em Kepler} cluster,
NGC~6791,  contain detached eclipsing binaries
\citep{Talamantes09,Street05,Marchi07,Mochejska05}. For these stars 
masses and radii can be determined independently \citep{Grundahl08}, which
will further strengthen results of asteroseismic analyses.

\section{Discussion \& Conclusions}\label{conclusions}
Photometric data of red giants in NGC~6819 obtained by NASA's {\it Kepler 
Mission} have enabled us to make the first clear detection of solar-like
oscillations in cluster stars. 
The general properties of the oscillations (\dnu, \numax, and amplitudes)
agree well with results of field red giants made by {\it Kepler}
\citep{Bedding10}  and CoRoT \citep{Ridder09,Hekker09}.
We find that the oscillation amplitudes of the observed stars scale as  
$(L/M)^{0.7}\,$\teff$^{-2}$, suggesting that previous
attempts to detect oscillations in clusters from ground were at the
limit of detection.

We find that the oscillation properties provide additional tests for
cluster membership, allowing us to identify four stars that are either
non-members or exotic stars.  All four stars have membership probability
higher than 80\% from radial-velocity measurements, but three of them
appear to be photometric non-members. 
We further point out that deviations from the theoretical predictions of
the asteroseismic parameters among a large sample of cluster stars 
have the potential of being used as
additional constraints in the isochrone fitting process, which
can lead to improved stellar models.  

Our results, based on limited data of about one month, 
highlight the unique potential of asteroseismology on
the brightest stars in the stellar clusters observed by {\it Kepler}.  
With longer series sampled at the spacecraft's short cadence ($\simeq$ 1 minute), we
expect to detect oscillations in the subgiants and turn-off stars, as well
as in the blue stragglers in this cluster.

\acknowledgments
Funding for this Discovery mission is provided by NASA's Science Mission Directorate.
The authors would like to thank the entire {\it Kepler} team without whom this
investigation would not have been possible.
The authors also thank all funding councils and agencies that have supported the activities of
Working Group~2 of the {\it Kepler} Asteroseismic Science Consortium (KASC).

{\it Facilities:}\facility{Kepler}.

\clearpage

\begin{table}
{\footnotesize
\begin{center}
\caption{Cross identifications and membership.\label{tab1}}
\begin{tabular}{lcccccc}
\tableline\tableline
ID        & ID    & WOCS ID   &  ID   & Mem.ship & Mem.ship & Mem.ship  \\
This work & KIC$^\mathrm{a}$ & Hole et al. & Sanders & Hole et al.$^\mathrm{b}$& Sanders$^\mathrm{c}$ & This work \\
\tableline
1    & 5024272 & 003003 &     & SM  95\% &      & no    \\
2    & 5024750 & 001004 & 141 & SM  93\% & 83\% & yes   \\
3    & 5023889 & 004014 & 42  & SM  95\% & 90\% & no    \\
4    & 5023732 & 005014 & 27  & SM  94\% & 90\% & yes   \\
5    & 5112950 & 003005 & 148 & SM  95\% & 92\% & yes   \\
6    & 5112387 & 003007 & 73  & SM  95\% & 88\% & yes   \\
7    & 5024512 & 003001 & 116 & SM  93\% & 90\% & yes   \\
8    & 4936335 & 007021 & 9   & SM  95\% & 68\% & no    \\
9   & 5024405 & 004001 & 100 & SM  93\% & 91\% & yes   \\
10   & 5112072 & 009010 & 39  & SM  95\% & 91\% & yes   \\
11   & 4937257 & 009015 & 144 & SM  88\% & 80\% & no    \\
\tableline
\end{tabular}
\tablenotetext{a}{ID from the {\it Kepler Input Catalogue} \citep{Latham05}.}
\tablenotetext{b}{Classification (SM: single member) and membership
  probability from radial velocity \citep{Hole09}.}
\tablenotetext{c}{Membership probability from proper motion \citep{Sanders72}.}
\end{center}}
\end{table}

\clearpage

\clearpage

\clearpage 

\end{document}